\documentclass[12pt]{book}

\usepackage[dvips]{graphicx,color}
\usepackage{makeidx,tsukuba}

\makeauthorindex
\makeindex

\begin{document}

\BookTitle{\itshape The 28th International Cosmic Ray Conference}
\CopyRight{\copyright 2003 by Universal Academy Press, Inc.}
\pagenumbering{arabic}

\chapter{Probing TeV gravity with extensive air-showers}

\author{Maximo Ave,$^1$ Eun-Joo Ahn,$^2$, Marco Cavagli\`{a},$^3$ and Angela
Olinto$^{1,2,4}$
\\
{\it (1) Enrico Fermi Institute, University of Chicago,
   5640 S.\ Ellis Avenue, Chicago, IL 60637, USA \\
(2) Dep. of Astron. \& Astrophys., Univ.
of Chicago,  Chicago, IL 60637, USA \\
(3) Institute of Cosmology and Gravitation, University of
Portsmouth, Portsmouth PO1 2EG, UK \\
(4) Center for Cosmological Physics, Univ.
of Chicago,  Chicago, IL 60637, USA} \\
}

\section*{Abstract}

Particle collisions with center-of-mass energy larger than the fundamental
gravitational scale can generate non perturbative gravitational objects such as
black holes and branes. In models with large extra dimensions,  the fundamental
gravitational scale may be around a TeV, making it possible for next generation
particle colliders and ultra-high energy cosmic rays to produce such non
perturbative gravitational objects. The decay of TeV gravitational objects is
significantly different from standard model processes such that 
probes of these new
ideas are within reach. We study the differences between standard model and TeV
gravity interactions in extensive air showers (EAS) generated by 
ultra-high energy
cosmic neutrinos. We show that discriminating TeV gravity from standard model
interactions is generally difficult, but not impossible given a few unique
signatures.

\section{Introduction}

In models with large extra dimensions (LEDs) [6-8], the fundamental scale of
gravity may be $\sim$ TeV. In these models, particle collisions with 
center-of-mass (CM) energy larger than  $\sim$ TeV may create non perturbative
gravitational objects such as black holes (BH) [9] and branes [2,3]. Next
generation particle colliders [11,13] and interactions of ultra-high energy
cosmic rays (UHECRs) with the atmosphere [12,5,4] can reach TeV CM energies
and, therefore, create these non perturbative gravitational objects (see [10]
for a complete review).

The shower-to-shower fluctuations in EASs initiated by hadronic primaries
combined with the small branching ratio to BH formation  makes the study of TeV
gravity with ultra-high energy protons hopeless [1]. Ultra-high  energy
neutrinos produced by the photo-pion production of ultra-high energy protons
in  the cosmic microwave background provide a cleaner beam to test departures
from SM interactions [17]. Here, we discuss the characteristics of EASs
initiated by cosmogenic neutrinos that produce BHs in TeV gravity theories and
contrast these with standard model (SM) interactions (based on a more detailed
study reported in [1]). We find that due to the long interaction length of high
energy neutrinos in the atmosphere and the large uncertainties in the  BH
formation cross section discriminating between the two scenarios is quite 
difficult. However, we also find that a few unique signatures involving tau
leptons can  help  detect TeV BHs.

\section{Black hole production in TeV gravity}

The 4-dimensional Planck mass in natural units can be written as $M_{Pl} \equiv
G_4^{-1/2}$, where $G_4$ is the 4-dimensional gravitational constant. In the
presence of extra dimensions, the fundamental Planck mass is given by
$M_{\star} \,=\, G_{n+4} ^{-1/(n+2)}$, where $n$ is the number of extra
dimensions. The 4-dimensional and $(n+4)$-dimensional gravitational constants
are related by $G_4 \,=\, {G_{n+4} / V_n} $, thus $M_{Pl}^2 \,=\,
M_{\star}^{n+2} ~ V_{n}$ , where $V_{n}$ is the volume of the extra dimensions.
In models where $V_n$ is large, $M_{\star}$ can be $\sim$ TeV. Depending on
details of LED models (such as $n$, $M_{\star}$, and minimum BH formation mass,
$M_{BH,min}$), the  cross section for BH formation in neutrino-nucleon
interactions can be either enhanced or suppressed with respect to SM  cross
section by several orders of magnitudes. Fig.~1 shows an example of the range
in  BH formation and SM cross sections, which include uncertainties in
$M_{BH,min}$  and the parton distribution functions (PDFs) [14]  (see [1] for
other examples). If  BHs  form in TeV CM collisions, they evolve by shedding
their hair first followed by Hawking evaporation, leading to the emission of SM
particles such as quarks  and gluons that subsequently hadronize into jets.
\begin{figure}[h]
   \begin{center}
     \includegraphics[height=7cm]{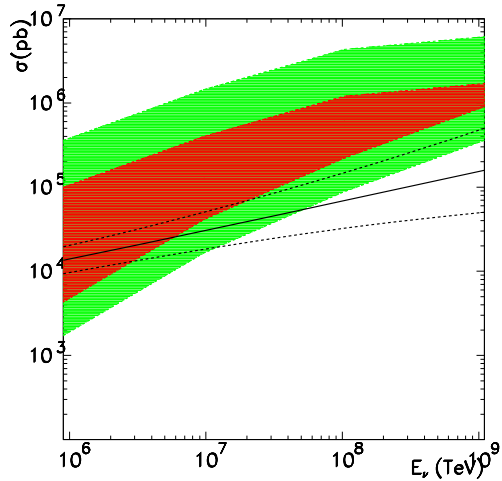}
   \end{center}
   \vspace{-0.7pc}
\caption{BH cross section for $n=6$ and $M_{\star}=1$ TeV, with the
$M_{BH,min}$ range in red and the uncertainties at the parton level and PDF in
green. The solid lines give the SM cross section, with dashed lines showing PDF
uncertainties for the SM case.}
\end{figure}

\section{Extensive air-shower simulations}

We simulated both SM showers and BH showers to compare their detectable
characteristics. The most relevant SM process for the comparison is the $\nu_e$
charged current (CC) interaction which produces an electron  shower that we
modeled with AIRES [15]. For the BH production process, the secondary particles
from BH evaporation are hadronized with PYTHIA [16] before producing a shower
with AIRES.
\begin{figure}[h]
   \begin{center}
     \includegraphics[height=6cm]{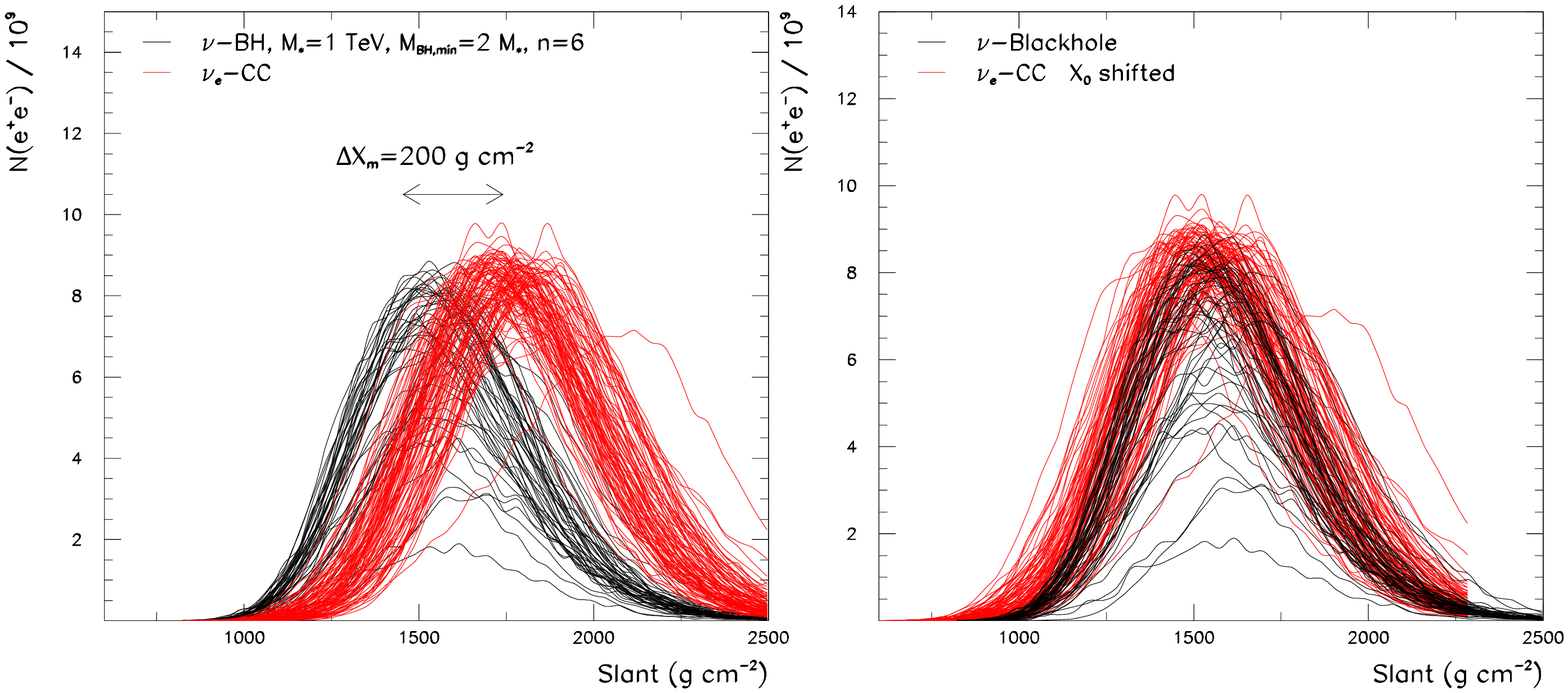}
   \end{center}
   \vspace{-0.5pc}
\caption{$e^+e^-$ pairs as a function of slant depth for $E_\nu=10^7$ TeV.
On the left,  $X_0^{CC}=X_0^{BH}$, while on the right,  $X_0$'s are 
shifted s.t.
$X_m^{CC} = X_m^{BH}$.}
\end{figure}
In Fig. 2, we show showers generated by neutrino primaries with energy
$E_\nu=10^7$ TeV and TeV gravity parameters  $n$=6, $M_{BH,min}=2M_{\star}=2$
TeV. On the left panel, the first interaction point ($X_0$) is fixed to be 10
km above sea level with shower zenith angle of $70^\circ$, corresponding to a
slant depth of 780 $g\,cm^{-2}$, for  both BH and SM showers
($X_0^{CC}=X_0^{BH}$). $\nu_e-$CC and BH showers are clearly  different by
$\sim 200$   g/cm$^{2}$. However, unlike the case of UHE protons, the 
interaction length for neutrinos in the atmosphere is large, thus, $X_0^{CC}$
is not fixed. By shifting $X_0$ such that the shower maxima, $X_m$, for both
cases match ($X_m^{CC} = X_m^{BH}$), the differences in shower development are
much harder to distinguish as seen in the right panel  of Fig. 2. Given a large
number of neutrino horizontal showers, one can distinguish the SM and the BH
cases by studying the rise of the shower, $X_m \,-\,X_{0.1}$, where $X_{0.1}$
is the slant depth containing 10 \% of  particles of $X_m$. In addition, the
muon content of BH showers is larger than the  SM case, since BHs produce
hadrons while CC showers do not. A deep horizontal shower accompanied by many
muon secondaries is a sign of BH formation.

Detecting TeV  BH formation with UHECR detectors may be possible  through the
decay of $\tau$-leptons generated by $\nu_{\tau}$'s that interact in the Earth 
or in mountain ranges close to the detectors. A secondary $\tau$ generated
through  the decay of a BH has much less energy than the SM $\tau$ secondary.
In addition,  BHs may produce multiple $\tau$-leptons in their evaporation, a
unique signature of  TeV gravity. SM processes that generate multiple 
$\tau$-leptons are highly unlikely, the detection of multiple $\tau$'s in
earth-skimming and mountain  crossing neutrinos will be a smoking gun for BH
formation.

\section{Conclusions}

We showed that given the uncertainties in the $\nu$-nucleon cross  section for
TeV gravity and the flux of cosmogenic neutrinos, distinguishing TeV LED 
models from the SM via the rate of neutrino induced EASs is unattainable.
Although BH  showers develop faster that the SM ones leading to a difference of
200 g/cm$^2$ in  $X_m-X_0$, the variation in $X_0$ for neutrino showers make
the distinction quite subtle. A large number of neutrino EASs with measured
muon content and $X_m\,-\,X_{0.1}$ is necessary for a clear distinction of SM
and BH formation. Finally, a  few background free signatures such as multiple
$\tau$'s and lower energy $\tau$ secondaries may more clearly signal the
existence of  TeV LED models. (We thank NSF and DOE for financial support.)

\section{References}


\re
1.\ Ahn,~E.~J., Cavagli\`{a},~M., and Olinto,~A.~V. 2003, Phys.Lett.B., 551, 1.
\re
2.\ Ahn,~E.~J. and Cavagli\`{a},~M. 2002, Gen.Rel.Grav., 34, 2037.
\re
3.\ Ahn,~E.~J., Ave, M., Cavagli\`{a},~M., and Olinto,~A.~V. 2003, submitted.
\re
4.\ Anchordoqui, et al., 2002,
Phys.Rev.D., 65, 124027.
\re
5. Anchordoqui, L. and ~Goldberg, H.,  2002, Phys.Rev.D, 65, 047502.
\re
6.\ Antoniadis,~I. 1990, Phys.Lett.B., 246, 377.
\re
7.\ Antoniadis,~I., et al.,  1998,
Phys.Lett.B., 436, 257.
\re
8.\ Arkani-Hamed,~N., Dimopoulos,~S., and Dvali,~G. 1998, Phys.Lett.B., 429,
263.
\re
9. Banks,~T. and Fischler,~W. 1999, hep-th/9906038.
\re
10. Cavagli\`{a},~M. 2003, Int.Jou.Mod.Phys.A., in press, hep-ph/0210296.
\re
11. Dimopoulos,~S. and Landsberg,~G. 2001, Phys.Rev.Lett, 87, 161602.
\re
12. Feng,~J.~L. and Shapere,~A.~D. 2002, Phys.Rev.Lett, 88, 021303.
\re
13. Giddings,~S. and Thomas,~T. 2002, Phys.Rev.D., 65, 056010.
\re
14. J.~Pumplin,et al. 2002,
JHEP., 0207, 012
\re
15. Sciutto, S.~J.~, 1999, astro-ph/9905185
\re
16. Sj\"{o}strand, T., et al., 2001, Computer Physics Commun., 135, 238.
\re
17. Tyler, C., Olinto, A.~V., and Sigl, G., 2001,
Phys.\ Rev.\ D {\bf 63}, 055001.

\endofpaper
\end{document}